\documentclass[prb]{revtex4}

\usepackage{graphicx}
\usepackage{dcolumn}
\usepackage{bm}

\begin{document}


\title{Research of the natural neutrino fluxes by use of large volume scintillation detector at Baksan}

\author{I.R. Barabanov, G.Ya. Novikova, V.V. Sinev, E.A. Yanovich}
\affiliation{%
Institute for Nuclear Research RAS, Moscow\\
}%


\begin{abstract}
A large volume scintillation detector is supposed to install at Baksan neutrino observatory INR RAS at Caucasus. 
The detector will register all possible neutrino fluxes, but mainly geo-neutrinos. In the article the neutrino fluxes 
are discussed and expected effects in proposed detector are estimated. The design and construction of such 
a detector is expected to be a part of all the world similar detectors net that is under construction in modern times.
\end{abstract}

\maketitle

\section*{Introduction}

It is known our days that natural neutrino fluxes carry out information as about the neutrino 
sources that can not be investigated directly as well about neutrino properties their selves. 

We consider here the research program for neutrino and antineutrino natural fluxes of low energy ($<$100$-$150 MeV) 
investigation by use of a large volume scintillation detector that we suppose to construct at Baksan neutrino 
observatory (BNO) INR RAS at a depth of 4800 m.w.e. The total target mass is supposed to be about 5 kt. 
In a number of papers there was demonstrated the possibility of using the detector like this for the considered goals [1$-$3].

The principal directions and goals of proposed researches connected with neutrino geo-physics and astrophysics:
\begin{itemize}
\item Searching for antineutrino flux emitted by uranium and thorium radioactive families decay products 
(geo-neutrinos) that are distributed in inner parts of the Earth and obtaining by this the radiogenic part of the Earth's thermal flux;
\item Testing of the hypothesis that there may a fission chain reaction exist in the centre of the Earth by looking for antineutrino flux from the "geo-reactor";
\item Research of the Supernovae blasts dynamics by detecting of the neutrino flash through the neutrinos energy spectrum and intensity;
\item Looking for isotropic neutrino flux coming from all gravitational collapses of massive stars nuclei that were taking place in 
the Universe during all its life continuing for billions years;
\item Detecting of the total antineutrino flux originating from all power nuclear reactors in the world. Confirming by this the neutrino 
oscillation parameters existing now;
\item Research of the solar neutrino spectrum jointly to other solar neutrino detectors.
\end{itemize}

Research of natural neutrino fluxes is under consideration for a long time.  But to measure them really was 
difficult because usually the detector background level was higher than a signal from neutrino flux. 
After the moment when new experimental setups KamLand in Japan [4] and BOREXINO in Italy [5] 
were began to work, it became clear that in experimental technique the revolution takes place. 
The inner background of a detector was decreased in hundreds times. So, it became possible to measure 
low intensity neutrino fluxes. The detector of large volume with liquid scintillator will be a multipurpose detector, 
it can detect all above mentioned neutrino fluxes. If inner detector background will be effectively suppressed 
there still be outcoming background connected with near passing high energy muons. That is why the so 
kind detectors must be placed as deep as possible under the earth.

Very popular now the idea of geoneutrinos detection is discussed very wide. It is connected with the problem 
that inner parts of the Earth could not be explored by other methods than neutrinos. Acoustic methods give 
only information on the density distribution inside the Earth [6], but neutrino registration method may give 
information about sources of radioactive elements inside the Earth and confirm or reject hypotheses on planet 
formation and on the elements behaviour at high pressures and temperatures. Also the detection of geoneutrino flux 
can make more precise the input of radioactive elements in the Earth's total thermal flux. For the moment they believe 
it is equal to 44$\pm$1 TW, but the estimated input of radioactive elements gives about a half of this value.

There is also a hypothesis of geo-reactor existence, which may take place in the centre of the Earth's core or on 
the boarder of the core and the mantle [7]. This hypothesis can explain all thermal flux, energy source of the 
Earth magnetic field and also the periodical change of magnetic poles of the Earth. It was shown that KamLAND 
detector can not measure antineutrino flux connected with geo-reactor. Proposed detector will be possible in several 
years confirm or reject this hypothesis.

The designing and construction of so kind a detector at Baksan neutrino observatory will be part of the all world 
detector net for geo-neutrino detection, that is under discussion in our days.

There are also two fundamental problems: registration of antineutrinos from Supernova and isotropic neutrino 
background flux from all gravitational collapses taking place in the Universe history. Discovering of the isotropic 
background permits to estimate the collapses rate and understand better the amount of matter hided from direct observation. 
In 1987 the Supernova was detected first time in the mankind history through the neutrino emission. It was occured in 
the Grand Magellan Cloud and received name SN1987A. The distance from the Earth was estimated as $\sim$50 kpc. 
Neutrino flux was registered by four active for that moment setups: Kamiokande, IMB, LSD and BNUT 
(Baksan neutrino underground telescope). The total statistics was only 24 events, registered during very short time interval. 
Detailed analysis of this detection is done in several reviews [8, 9].

Now there are several detectors are included in the world wide net for looking for neutrino bursts from Supernovae. 
Each of them has advantages and disadvantages.

Proposed detector gives bigger opportunities in neutrino detection from Supernovae. According to our estimations in 
this detector we could register about 100 events if the similar SN1987A neutrino burst took place. At distances closer 
than 10 kpc the spectral analysis becomes possible, this could give the information on neutrino mass hierarchy and get 
better oscillation parameters [3]. At a distance 10 kpc the detector with 5 kt target can register about 1500 antineutrino 
events. By the way proposed detector can register and other neutrino reactions (interactions of neutrinos with nuclei of 
carbon and neutrino scattering on carbon nuclei and free electrons), what could give better reliability in detection of SN.

The detector could also register neutrino flux from the Sun if the appropriate scintillator purity will be achieved 
(at the same level as BOREXINO one [5] $\sim$10$^{-18}$ g/g). The standard neutrino registration reaction is supposed to 
be used - the weak scattering of neutrinos on electrons. If the target mass will be factor 20 bigger than the BOREXINO 
one, the better accuracy may be achieved in registration $^{7}$Be and CNO cycle neutrinos.

\section{Geoneutrinos}

To explain the heat flux outgoing from the inner parts of the Earth there are several hypotheses. One of them is the 
existence of radioactive elements in the innermost part of the Earth which produce heat by means of the energy 
released in alpha- and beta-decays. During beta-decay the antineutrinos are emitted, and they go away from the 
Earth and can be detected near the Earth surface.  This antineutrino emission is called geoneutrinos.

They regard geoneutrinos to be the antineutrinos produced during a set of radioactive nuclei having half-life periods 
compared with the Earth age and located at the Earth inner part. They are $^{40}$K, $^{238}$U and $^{232}$Th.  There are some 
other isotopes such as $^{235}$U and $^{87}$Rb, but their input in total geoneutrino flux is negligible and usually not considered.

On the basis of seismic data it is known that the Earth consists of several spherical layers. The outside layer is a crust 
which consists of tectonic plates and thin oceanic crust. The crust thickness is estimated about 5-6 km at ocean bottom 
up to 30$-$60 km  in the region of tectonic plates. Sometimes in the models the crust is described as spherical layer 30$-$50 km 
wide. After the crust they regard upper mantle to spread down about 550$-$570 km from the surface. The lower mantle goes 
deeper, down to 2900 km, and is separated from the upper one by the thin peridot layer. Deeper than the mantle, there is a 
nucleon or the core which is separated from the outer - liquid (2900$-$5150 km) and central $-$ solid (deeper than 5150 km).

According to modern views all radioactive isotopes are located in the crust and in the mantle in equal proportions.  
The core does not contain radioactive elements at all according to the geochemists opinion. 

In a detector placed under the surface of the earth mainly the antineutrinos from the crust are to be detected. 
But the total counting rate will be different depending on the detector location. It may be placed in the point with 
greater or smaller crust width. The estimations [10$-$12] show that counting rate at Hawaii will be about 13 TNU (1 TNU = 1 
event/10$^{32}$ target protons per year), but at Baksan about 55 TNU.

The geoneutrino measurements are just carried out. Collaboration KamLAND (Kamioka, Japan) [4] began the measurements in 
2002, and in 2005 they reported the first observation of geoneutrinos. They published the result: 28 $^{+16}_{-15}$ events that 
correspond to 57 $^{+33}_{-31}$ TNU [10] in accordance with predictions of BSE model. Main limitations come from a 
large non-removable background coming from working nuclear power plants, see table 1. For the BOREXINO detector 
(Gran Sasso, Italy) the same background is comparable with an effect, but the effect itself is low enough (5$-$7 events/year) [14, 15].

For realizing the item of the Earth thermal flux, searching for the $^{238}$U and $^{232}$Th inputs in the total heat production, containment of 
uranium and thorium in the crust and mantle etc., it demands reliable statistics and spectral analysis of events. That is why foreseeing 
progress in research of geoneutrinos is linked to new detectors similar or larger than KamLAND, but located far from nuclear power plants. 

Such a detector is proposed to be constructed at Baksan neutrino observatory RAS, which could have counting rate for 
geoneutrinos at the level of 220 events per year. The measured positron spectrum from inverse beta-decay reaction of geoneutrinos is shown at figure 1
\begin{equation}
\bar{\nu_{e}}+p \rightarrow n + e^{+}.
\end{equation}

\begin{figure}
\includegraphics{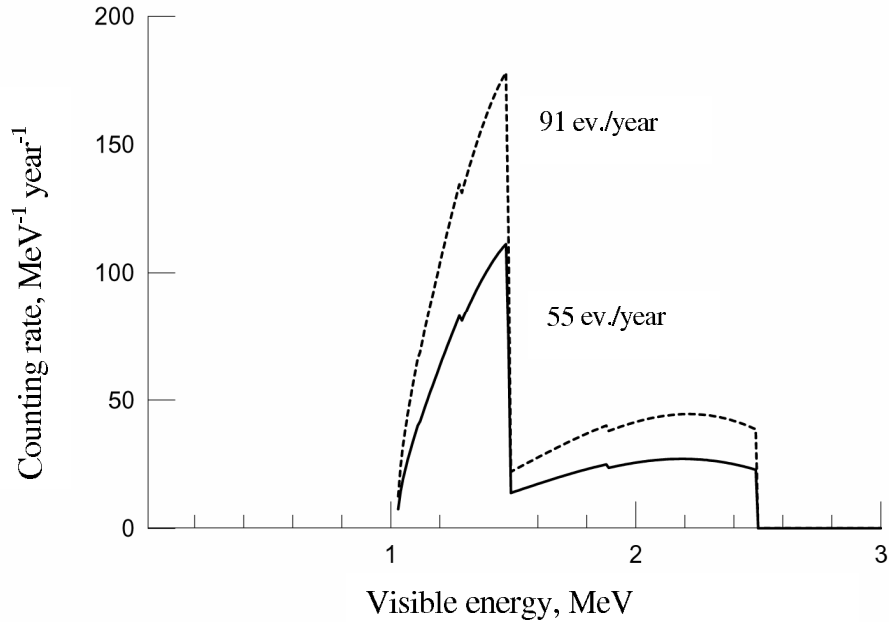}
\caption{\label{fig:fig1} Positron spectrum produced in the detector by geoneutrinos. Dashed line is for the case of oscillation absence. 
The spectrum is calculated for antineutrino flux from [10] for 10$^{32}$ target protons.}
\end{figure}

\begin{table}[h]
\caption{Expected counting rates $N_{geo}$ events of reaction (1) in some sites on the Earth in total geoneutrino 
flux from decays of U and Th (accounting oscillations) in TNU (1 TNU = 1 event per year in a target with 10$^{32}$ protons). 
$R$ is the ratio of reactor flux rate $N_{reactor}$ to geoneutrino one $N_{geo}$ in geoneutrino energy area.}
\label{table:1}
\vspace{10pt}
\begin{tabular}{l|c|c|c|c|c}
\hline
Location &Depth, mwe & \multicolumn{3}{c|}{Counting Rate $N_{geo}$, TNU} & $R=\frac{N_{reactor}}{N_{geo}}$ \\
\cline{3-5}
 &  & [10] & [11] & [12] &  \\
\hline
Hawaii (USA)& $\sim$4000 & 13.4 & 13.4 & 12.5 & 0.1 [10] \\
Kamioka (Japan)& 2700 & 36.5 & 31.6 & 34.8  & 6.7 [10] \\
Gran Sasso (Italy)& 3700 & 43.1 & 40.5 & 40.5  & 0.9 [14, 15] \\
Sudbury (Canada)& 6000 & 50.4 & 47.9 & 49.6  & 1.1 [10] \\
Pyhäsalmy (Finland)& 4000 & 52.4 & 49.9 & 52.4  & 0.5 [10] \\
Baksan (Russia)& 4800 & 55.0 & 50.7 & 51.9  & 0.2 [1, 46] \\
\hline
\end{tabular}\\[2pt]
\end{table} 

\section{Georeactor}

The existence of radioactive elements in the Earth does not explain the total thermal flux of the Earth. 
As well it can not explain the Earth's magnetic field source and periodical change of magnetic poles. J. M. 
Herndon [6] proposed a hypothesis of natural nuclear reactor existing in the center of the Earth to explain these effects. 
This nuclear reactor, he called it georeactor, is to have power of 3$-$10 TW to describe the necessary part of heat flux 
and intensity of observed magnetic field. This hypothesis explains by the way the periodical change of magnetic poles 
with increasing and decreasing of the Earth magnetic field. Georeactor is periodically poisoned by its decay products 
and stops, but than after these decay products are removed from ``active zone" it starts working further.

The hypothesis could be tested with the future Baksan detector. The counting rate of the detector because of 
georector antineutrino flux is estimated as 80$-$250 events per year at 100\% registration efficiency and 
accounting neutrino oscillations, see figure 2. The spectrum shape does not change affecting oscillations 
because of far distance from the reactor ($\sim$6000 km), but intensity will fall down the coefficient 0.59.

So, during one year measurements the hypothesis of georector in 3 or more TW could be confirmed or rejected.

\begin{figure}
\includegraphics{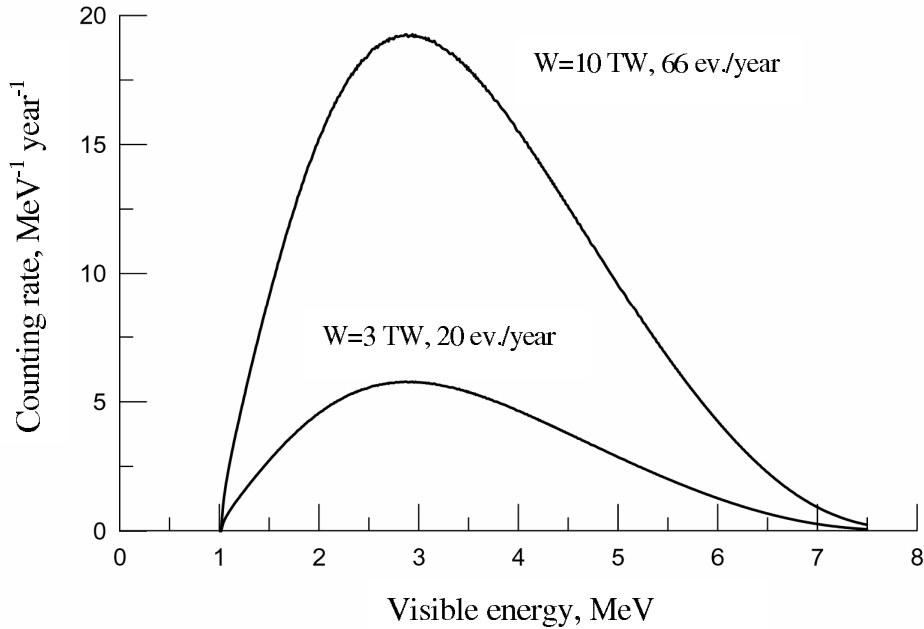}
\caption{\label{fig:fig2} Positron spectra observed in the detector from georeactor antineutrino flux. 
Spectra are shown accounting oscillations and calculated for 10$^{32}$ target protons.}
\end{figure}

\section{Power nuclear reactors background}

The detector would register as background the antineutrino flux from all nuclear power plants built 
all over the world. But this background can be taken into account and calculated with accuracy $\sim$3\% [1].

At figure 3 one can see standard positron spectrum from nuclear reactors registered by the detector. 
The curve 2 is the spectrum affected by oscillations, it has specific peaks depending from mutual positions 
of the detector and reactors. The number and peak positions on spectrum are described by oscillation parameters. 
Figure 4 shows the difference of this background from some possible places of detector location.

Estimated counting rate of nuclear reactors flux for Baksan is $\sim$160 events per year in 5 kt target. 

\begin{figure}
\includegraphics{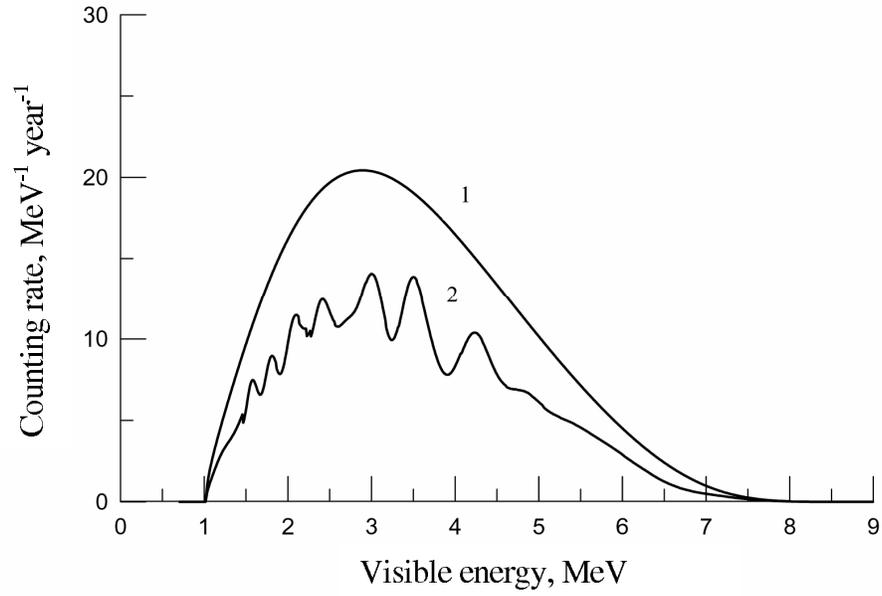}
\caption{\label{fig:fig3} Positron spectra observed in the detector from nuclear reactors antineutrino flux. 
Curve 1 is for no oscillations case ($\sim$70 ev./year) and line 2 includes oscillation effect ($\sim$40 ev./year). 
Shown spectra are calculated for 10$^{32}$ target protons.}
\end{figure}

\begin{figure}
\includegraphics{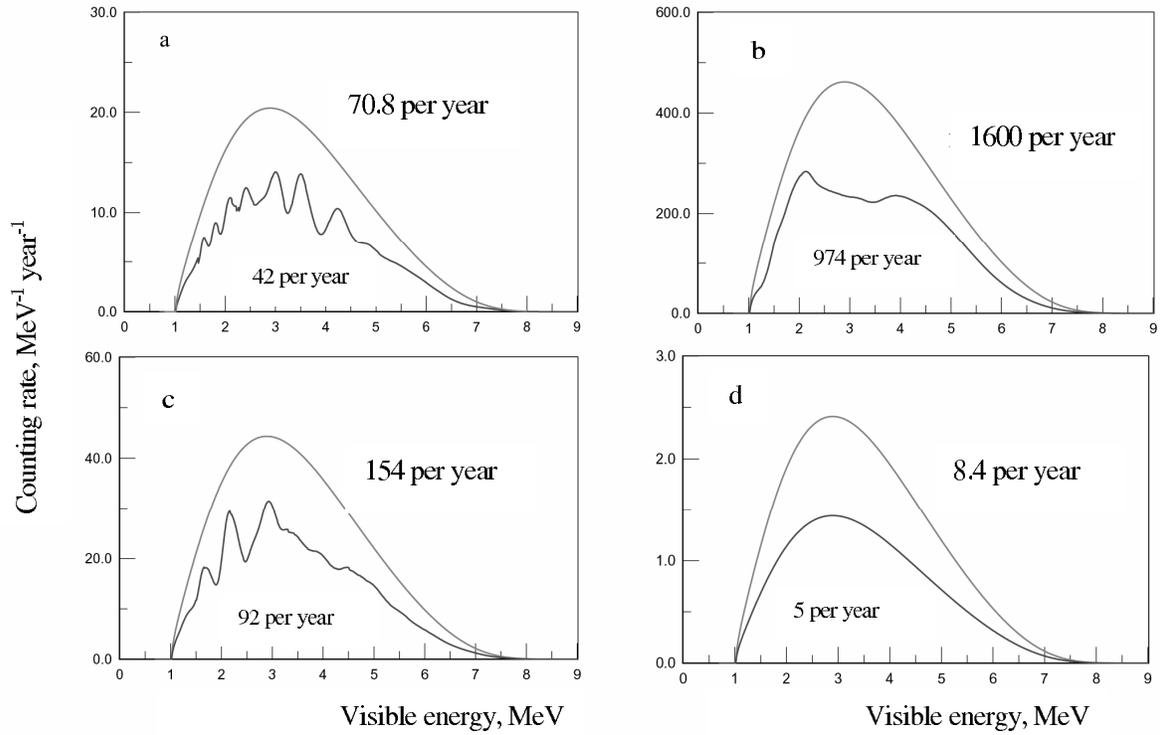}
\caption{\label{fig:fig4} Positron spectra similar to shown at figure 3 but for several positions of proposed detector: ? - Baksan, b - KamLAND, c - Pyhasalmy, d - Hawaii.}
\end{figure}

\section{Supernovae}

``Thermal'' neutrino flash begins simultaneously with gravitational collapse of massive stars iron core ($M>8M_{sol}$) 
and it continues about 20 s. All 6 types of active neutrino flavors are produced in burning hot core: $\nu_e$, $\bar{\nu}_e$, $\nu_{\mu}$, $\bar{\nu}_{\mu}$ and 
$\nu_{\tau}$, $\bar{\nu}_{\tau}$. They carry out the overwhelming part of gravitational energy $\sim 3\cdot 10^{53}$ erg, 
released in a collapse. Only after ten minutes or several hours the blast achieves the star surface, and Supernova 
lights up on the sky and one can observe it with the naked eye and with the use of radio, optical, X-ray and 
gamma astronomy methods. Afterwards the largest part of the star substance dissipates in the Cosmos, 
but the neutron star (or a black hole) appears instead of the star.

To the moment of the neutrino flash beginning the inner part of a core becomes opaque for neutrinos because of high 
density. The neutrinos are multiply scattered, absorbed and emitted once more before outgoing from the core. 
The surface from which neutrinos can escape the star is called neutrino sphere. The radii of neutrino spheres are 
different for neutrino flavours because of different interaction cross sections. The deepest one is neutrino sphere for
$\nu_{\mu}$, $\bar{\nu}_{\mu}$ and $\nu_{\tau}$, $\bar{\nu}_{\tau}$ (radius $\sim$30 km), further is neutrino sphere for
$\bar{\nu}_e$ lays ($\sim$50 km), and the last one is neutrino sphere for  $\nu_e$ ($\sim$70 km). This fact explains 
the difference in mean energies of neutrino escaping from the core. The larger radius the less is mean energy. 
However, as calculations show, all neutrino flavours share carried out energy in about equal proportions. 

Directly before ``thermal" flash the short electron neutrino $\nu_e$ pulse ($\sim$10$^{-2}$ s) appears, that origins from the neutronization 
process of a core, coming to the limit of stability (Chandrasekhar limit).These $\bar{\nu}_e$'s, having energies 15$-$20 MeV, 
carry out 5$-$10\% of the energy released in the collapse.

Thus, the researches of the neutrino flash is a powerful tool to search for collapsing star energetic and dynamics as well as neutrino properties themselves.

Firstly it was designated to the existence of the neutrino flash, corresponding to the collapse in 1965 [16]. 
In resent years the intensive calculations of Supernova blast dynamics and accompanying neutrino emission were done [17$-$20]. 
One can see detailed reviews (see for example [21, 22] and citations therein). The method of neutrino flash registration was also 
proposed in 1965 [23]: one can observe statistically significant series of signals in massive enough low background detector. However 
the Supernova blast is rare enough event (in our Galaxy we expect one blast per $\sim$30 years). That is why the authors 
[23] proposed to synchronize the operation of different detectors for better reliability and information. In our days there exists 
international system SNEWS [24, 25] (SuperNova Early Warning System), which unites all detectors having capability to 
detect neutrinos from Supernovae. One of the goals for such kind of systems is to send the message for astronomers about the 
expected Supernova appearance.

Neutrinos produced during the collapse play significant role in nucleosynthesis [26$-$28]. Observed prevalence of some light 
elements ( $^9$Be, $^{11}$B, $^{19}$F etc..) as well as prevalence of so called passed isotopes, that are heavier 
than iron, is possible to explain with the help of neutrino reactions taking place at the radius of $\sim$1000 km from the star core.

Let us note that neutrino flash registration is an oscillation experiment with very long base $\sim$30 thousand light years and 
produced while star neutrinos pass star regions where density exceeds the density in the centre of the Sun in large number of orders. 
Flavour composition of neutrinos coming to the detector depends on the value of mixing parameter $\sin^2{\theta}_{13}$
and on the mass hierarchy type (normal or inversed) [29, 30]. This fact reveals principally new abilities for searching for mixing 
and mass structure of neutrino that can exceed in sensibility experiments at reactors and accelerators. The transformations of
$\bar{\nu}_{\mu}, \bar{\nu}_{\tau}\rightarrow \bar{\nu}_{e}$ allow to measure intensities and energies of  
$\bar{\nu}_{\mu}$ and $\bar{\nu}_{\tau}$, signals which can not be measured directly with the modern methods.

Neutrino and antineutrino will produce several effects in the detector. The largest number of events ($\sim$1200, see table 2) 
is expected from the reaction (1). But using as a target carbon nuclei $^{12}$C will allow to divide ${\nu}_{e}$ and $\bar{\nu}_{e}$
according to the characteristic reaction products and decays of daughters isotopes $^{12}$B and $^{12}$N:
\begin{equation}
\bar{\nu_{e}}+^{12}{\rm C} \rightarrow ^{12}{\rm B} + e^{+} \ (E_{th}=13.9 \ {\rm MeV}),\  \ ^{12}{\rm B}\rightarrow ^{12}{\rm C}+ e^-
\ ({\tau}_{1/2}=20.2 \ {\rm ms}),
\end{equation}
\begin{equation}
\nu_e+^{12}{\rm C} \rightarrow ^{12}{\rm N} + e^{-} \ (E_{th}=17.8 \ {\rm MeV}),\  \ ^{12}{\rm N}\rightarrow ^{12}{\rm C}+ e^+
\ ({\tau}_{1/2}=11 \ {\rm ms}),
\end{equation}

Reactions (2, 3) have extremely high thresholds that's why the number of events will be very low in the absence of oscillations.

All neutrino flavors $\nu_i$ can elastically scatter on $^{12}$C nuclei
\begin{equation}
\nu_i + ^{12}{\rm C} \rightarrow ^{12}{\rm C} + \gamma + \nu^{'}_{i} \ (E_{th}=15.1 \ {\rm MeV}) 
\end{equation}
as a result of the monochromatic line of single gammas will appear and be registered with a detector. Because of low energies $\nu_e$
and $\bar{\nu_{e}}$ their summed input in total effect of reaction (4) does not exceed 5\% and the dominated role will be played by hard 
$\nu_x$ = (${\nu}_{\mu}, \bar{\nu}_{\mu}, {\nu}_{\tau}, \bar{\nu}_{\tau}$) neutrinos. The cross sections of reactions (2$-$4) were taken from [31]. 

Also all neutrino flavors can take part in the reaction of elastic scattering on free electrons
\begin{equation}
\nu_i + e^{-}\rightarrow e^{-'}+ \nu^{'}_{i}
\end{equation}
and registration of recoil electrons can give an opportunity to find principally the direction on supernova.

The observed in a detector positron spectrum will depend on the character of neutrino oscillations. Some possible variants are shown at figure 5.

\begin{figure}
\includegraphics{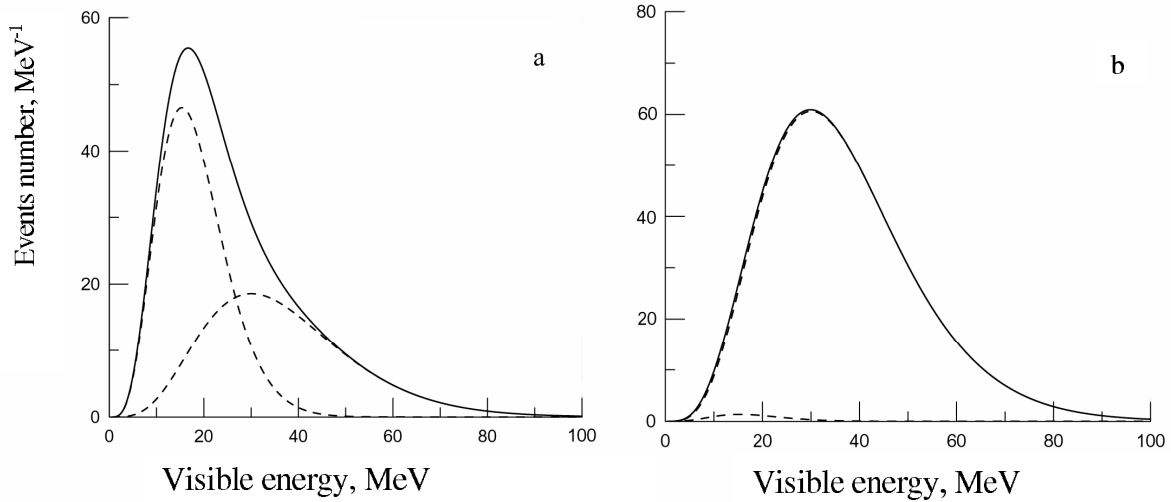}
\caption{\label{fig:fig5} The inverse beta-decay reaction positron spectra from SN accounting oscillations: ? - normal mass hierarchy. 
Dashed line shows the partial spectra of (muon and tau antineutrinos are more hard), b - the inverted mass hierarchy, $\sin^2{\theta}_{13}>10^{-3}$.}
\end{figure}

\begin{table}[h]
\caption{Expected number of neutrino events in scintillation detector with a target mass 5 kt (4$\cdot10^{32}$ protons, 16$\cdot10^{32}$ electrons,
2$\cdot10^{32}$ $^{12}$C nuclei) produced by the ``thermal" neutrino flash during the collapse of the massive star core (the distance to 
SN is 10 kiloparsec, the energy taken away by neutrinos is $\sim3\cdot10^{53}$ erg, normal mass hierarchy).}
\label{table:2}
\vspace{10pt}
\begin{tabular}{l|c|c|c}
\hline
Reaction & No oscillations & LMA MSW & LMA MSW \\
              &                        & ($\sin^2{\theta}>10^{-3}$) & ($\sin^2{\theta}<10^{-5}$) \\
\hline
$\bar{\nu_{e}}+p \rightarrow n + e^{+}$ & 1157 & 1479 & 1479   \\
\hline
$\bar{\nu_{e}}+^{12}{\rm C} \rightarrow ^{12}{\rm B} + e^{+}$ & 14.4 & 35.5 & 35.5  \\
${\nu_{e}}+^{12}{\rm C} \rightarrow ^{12}{\rm N} + e^{-}$ & 5.8 & 132 & 93.5  \\
\hline
$\sum ^{12}{\rm C}(\nu_{i},\nu^{'}_{i})^{12}{\rm C}+\gamma \ ^{*}$ & 236 & 236 & 236  \\
\hline
$\sum \nu_i + e^{-}\rightarrow e^{-'}+ \nu^{'}_{i}$ & 70.6 & 62.2 & 61.4  \\
\hline
\end{tabular}\\[2pt]
$^{*}\nu_i=\nu_{e}, \bar{\nu_{e}}, \nu_{\mu}, \bar{\nu_{\mu}}, \nu_{\tau}, \bar{\nu_{\tau}}$
\end{table} 

\section{Isotropic background from previous stellar gravitational collapses}

Relic neutrinos produced during previous collapses of massive stellar cores contain all flavours of neutrinos. But for the 
present moment while they use inverse beta-decay reaction for registration only electron antineutrinos can be detected. 
Expected total flux of SN relic neutrinos (SRN) is very small and according to different estimations equals to 
12$-$16 cm$^{-2}$ s$^{-1}$ [32]. The discovery of this flux can give some new information concerning gravitational 
collapses and neutrino properties (masses, mixing, magnetic moment). This flux carries out information concerning the 
collapses rate in the Universe [32$-$36]. The probability of detecting SRN is high enough for proposed detector 
because of the best limitation $\sim$1.4$-$1.9 $\bar{\nu_{e}}$ cm$^{-2}$ s$^{-1}$ for $E_{\bar{\nu_e}}>$ 19.3
MeV (90\% CL),
that was received by Super-Kamiokande [36] detector is only several times higher than theoretical estimations of the flux 
SRN for this energy region. Expected rate of SRN for the LENA project [38] (the target volume 2.9$\cdot10^{33}$ protons [35]) 
if to place it in the Centre of underground physics at Pyhäsalmy (Finland), is $\sim$4$-$6 events per year for energy range 
$E_{\bar{\nu_e}}=$ 9.7$-$25 MeV, see [35, 36, 38]. Energy area is limited by the background of reactor antineutrinos in 
soft part of SRN spectrum and atmospheric neutrinos in hard area.

At figure 6 all expected spectra from inverse beta-decay reaction (1) are shown including SRN (curve 5). 
The data for SRN were recalculated according to the data presented for LENA detector [35]. 
We took the same suggestions (direct mass hierarchy, $\bar{\nu_e}$ SRN spectrum from, [32] with normalization parameter
$f_{{\rm SN}}$ = 1 for LL supernova model). Reactor antineutrino rate was calculated with the use of spectrum [39]. 
Reactor antineutrino flux and atmospheric one at Baksan site are a factor of 2 less than at Pyhäsalmy (see table 1). 
This admits to use wider energetic range for measuring relic neutrinos, see figure 6. In the range of 
$E_{\bar{\nu_e}}=$ 8.0$-$30
MeV expected rate of SRN may be about 1 per year.

\begin{figure}
\includegraphics{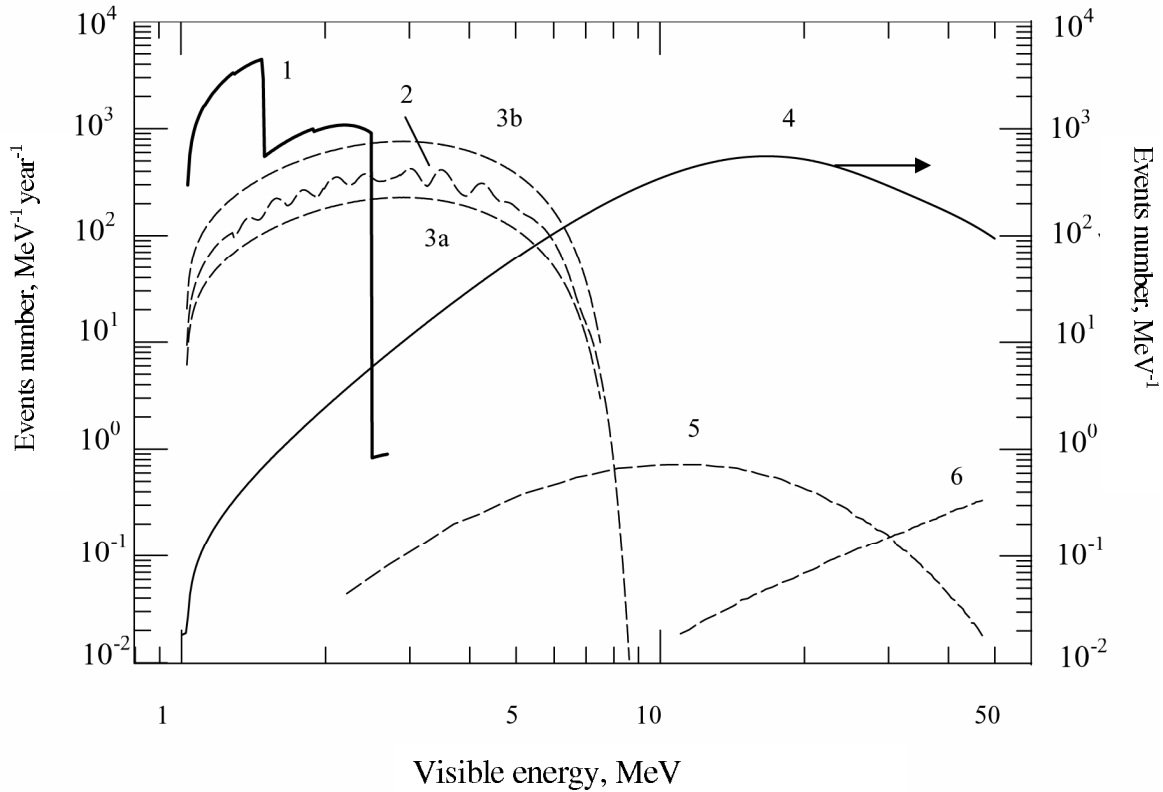}
\caption{\label{fig:fig6} Positron spectra from reaction (1) in Baksan detector (target is organic scintillator 5 kt, 
oscillations included): 1 $-$ geoneutrino, 2 $-$ $\bar{\nu_{e}}$ from nuclear power plants, 3a, 3b $-$ $\bar{\nu_{e}}$
from hypothetical georeactor, 4 $-$ $\bar{\nu_{e}}$ from supernova, see text (right scale), 5 - relic
$\bar{\nu_{e}}$ from gravitational collapses, 6 - atmospheric $\bar{\nu_{e}}$.}
\end{figure}

\section{Solar neutrinos}

The Sun is powerful enough source of electron neutrinos. Thermonuclear reactions providing solar burning energy produce 
neutrinos as well. The reactions producing neutrinos are the next:
\begin{eqnarray}
p+p \rightarrow ^{2}{\rm H} + e^{+}+ \nu_{e} \ \ 99.96\% \ (E_{\nu}<0.423 \ {\rm MeV})  \nonumber \\
p+e^{-}+p \rightarrow ^{2}{\rm H} + \nu_{e} \ \ 0.44\% \ (E_{\nu}=1.445 \ {\rm MeV})  \nonumber \\
^{7}{\rm Be}+e^{-} \rightarrow ^{7}{\rm Li} + \nu_{e} \ \  (90\% \ E_{\nu}=0.863 \ {\rm MeV}, 10\% \ E_{\nu}=0.385 \ {\rm MeV})  \nonumber \\
^{7}{\rm Be}+p \rightarrow ^{8}{\rm B} + \gamma , \  ^{8}{\rm B} \rightarrow 2\alpha + e^{+}+ \nu_{e} \ \  (E_{\nu}<15 \ {\rm MeV}) \nonumber \\
^{3}{\rm He}+p \rightarrow ^{4}{\rm He} +e^{+} + \nu_{e} \ \  (E_{\nu}<18.8 \ {\rm MeV})
\end{eqnarray}

The last reaction has very low intensity (three powers less than boron neutrinos) and may be omitted at first approximation. 
At figure 7 all solar neutrino spectra are shown according to the calculations made by Bahcall and Serenelli [40].

\begin{figure}
\includegraphics{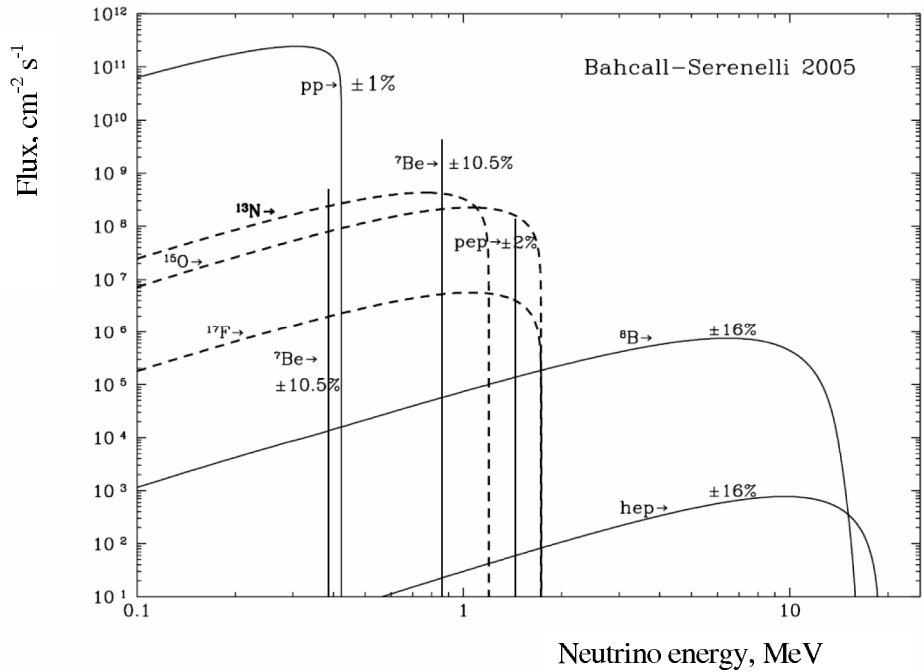}
\caption{\label{fig:fig7} Solar neutrinos energy spectra [40]. Dashed line shows spectra form CNO cycle.}
\end{figure}

Solar neutrinos have been registered during more than half of a century. First experiment $^{37}$Cl$-^{37}$Ar [41] 
had threshold 0.81 MeV and detected in general only boron and pep neutrinos. The Ga$-$Ge experiment [42] has 
smaller threshold 233 keV and includes in registration $^{7}$Be and partially pp neutrinos. The experiment only very 
recently added to setups registering solar neutrinos BOREXINO [5] having the lowest threshold for the moment $\sim$200 keV 
presented first results of measuring the neutrino flux from $^{7}$Be. Looking forward this experiment is going to distinguish the pep 
and CNO fluxes and possibly pp [43]. The SNO experiment (Canada) [44] having as a target heavy water had the threshold 
1.4 MeV for charge channel and 2.2 MeV for neutral channel of neutrino interaction with deuteron. 
This experiment firstly demonstrated the accordance of measured solar neutrino flux with theoretically predicted one.

Proposed detector will register recoil electrons from the neutrino-electron scattering reaction. At figure 8 one can see recoil 
electrons spectrum that can be measured in a detector as a result from the Sun neutrino flux.

\begin{figure}
\includegraphics{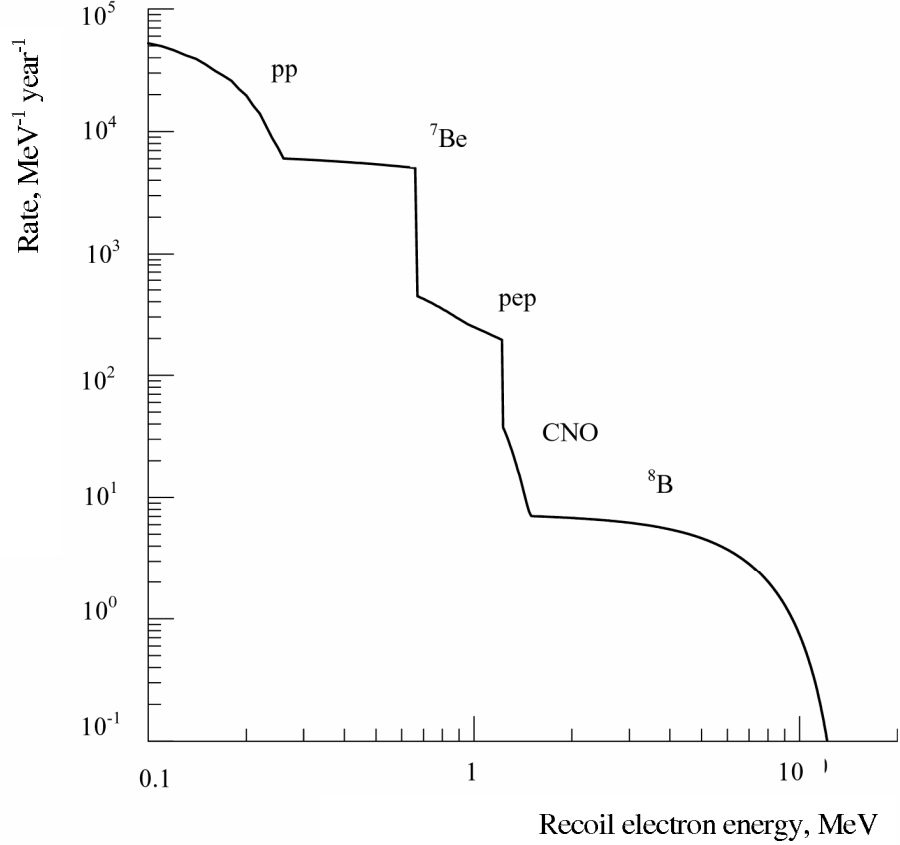}
\caption{\label{fig:fig8} Recoil electron energy spectrum from solar neutrino scattering reaction on the target electrons 
($16\cdot10^{32}$ electrons at 5 kt target). In calculated spectrum oscillations are not taken in to account.}
\end{figure}

\section{Neutrino detector}

Baksan neutrino detector would contain three concentric spherical zones. Central zone is a target with a diameter 
22.5 m and will be filled with organic scintillator. The second one having width of 2 m and filled with non scintillating 
liquid (mineral oil) serves for protecting a target from radioactivity of PMTs. It is separated from a target by transparent 
strong film (nylon type) or by Plexiglas. PMTs will be installed on the surface of the second zone made of stainless steel. 
Third zone is veto zone. It is filled with pure water or mineral oil and scanned by PMTs, which detect the Cerenkov 
light from passing muons. External dimensions of the detector might be about 33 m.

The organic scintillator we suppose to use will be made on the basis of pure LAB (Linear alkyl benzene) or its mixtures 
with some other solvents like PC, PXE etc. Similar scintillator is under consideration in other collaborations in particular 
SNO+ [45]. Scintillator should have high light yield and proper transparency to provide high energetic resolution. 
While constructing the detector of so considerable dimensions the very important question will be concerning safety 
of the scintillator and its expenditure. Using of LAB seems to satisfy these conditions. LAB is produced in high quantities 
and has low price. The flash point of LAB is 130C. 

Detector will register light pulses correlated in space and time produced by positron and neutron from reaction (1) 
captured in the target. Neutron is captured by hydrogen or Gd, which could be added to scintillator for increasing the 
efficiency of neutron registration. Mean time of neutron life before capture is $\sim$200 ms in the absence of Gd or 
30$-$50 $\mu$s at $\sim 1-$0.5 g/l Gd concentration. The useful target volume contains $4\cdot10^{32}$ protons 
(hydrogen nuclei), $2\cdot10^{32}$, $^{12}$C nuclei and $16\cdot10^{32}$ electrons.

In its structure the detector will be close to the detector KamLAND [4] and differs from it in larger target mass and 
deeper position under the ground. Detector scheme is shown at figure 9.

\begin{figure}
\includegraphics{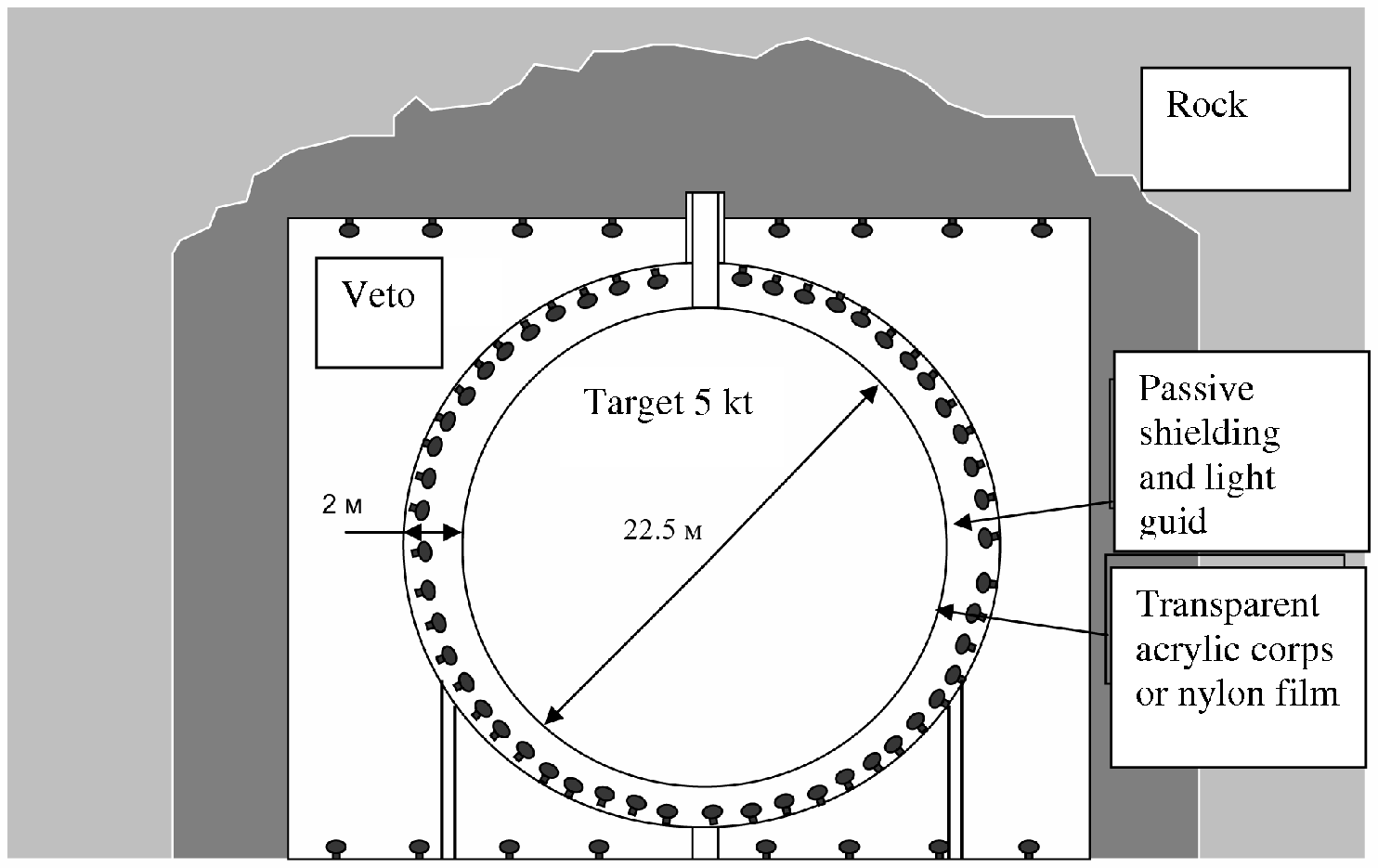}
\caption{\label{fig:fig9} Neutrino detector construction.}
\end{figure}

\section*{Conclusion}

The modern strategy in development of neutrino astro- and geo-physics is constructing of underground neutrino 
observatory net containing multipurpose experimental set-ups. In the article the program of searching for natural 
neutrino fluxes using the large scintillation detector with target mass of 5 kt is considered. It is shown that high 
statistics in combination with low background admit to make the next step in a progress of low energy neutrino physics.

Authors gratefully thank G. V. Domogatsky and L. A. Mikaelyan for the interest in the project and useful discussions, 
V.N. Gavrin and all colleagues from Baksan neutrino observatory INR RAS for hospitality and friendly discussions.

\end{document}